# Ricci Curvature of the Internet Topology

Chien-Chun Ni*, Yu-Yao Lin*, Jie Gao*, Xianfeng David Gu* and Emil Saucan†

*Department of Computer Science, Stony Brook University. {chni, yuylin, jgao, gu}@cs.stonybrook.edu
†Department of Mathematics, Technion, Israel Institute of Technology. semil@tx.technion.ac.il

***Abstract*—Analysis of Internet topologies has shown that the Internet topology has negative curvature, measured by Gromov's "thin triangle condition", which is tightly related to core congestion and route reliability. In this work we analyze the discrete Ricci curvature of the Internet, defined by Ollivier [1], Lin *et al.* [2], etc. Ricci curvature measures whether local distances diverge or converge. It is a more local measure which allows us to understand the distribution of curvatures in the network. We show by various Internet data sets that the distribution of Ricci cuvature is spread out, suggesting the network topology to be non-homogenous. We also show that the Ricci curvature has interesting connections to both local measures such as node degree and clustering coefficient, global measures such as betweenness centrality and network connectivity, as well as auxilary attributes such as geographical distances. These observations add to the richness of geometric structures in complex network theory.**

## I. Introduction

The Internet topology has been a popular subject of study for the past twenty years, at three different granularities: router level topology, the autonomous systems (AS) level topology, and the application level overlay topology. Understanding the Internet topology at all three levels are important. Such structural discoveries could help us better understand important system properties such as robustness and vulnerability, as well as how information flows on such topologies in terms of congestion control and virus diffusion. In the past few years there have been a number of interesting discoveries on the graph and geometric properties of the Internet topologies that hint for some deeper theory underneath.

### A. Prior Work

**Power Law Degree Distribution.** Faloutsos *et al.* [3], [4] studied the AS-level topology [5] and observed that the degree distribution follows a power law. That is, the number of nodes having degree $k$ is proportional to $1/k^{\gamma}$, $\gamma$ is between 2 and 3. A detailed description [6] says that the Internet topology looks like a "jellyfish", with a core in the middle and many tendrils connected to it. This is confirmed in data sets collected in later efforts [7], [8]. For the router level topology, an interesting dataset was obtained in the Rocketfuel project [9]. The outdegree was also found to have large variations, though the range is smaller compared to the AS-levels, due to physical constraints of a router. There were also studies of network topologies from peer-to-peer network topology (Gnutella [10], BitTorrent network [11]) and the Pretty Good Privacy (PGP) web of trust [12]. Both of them were also shown to have power law degree distribution.

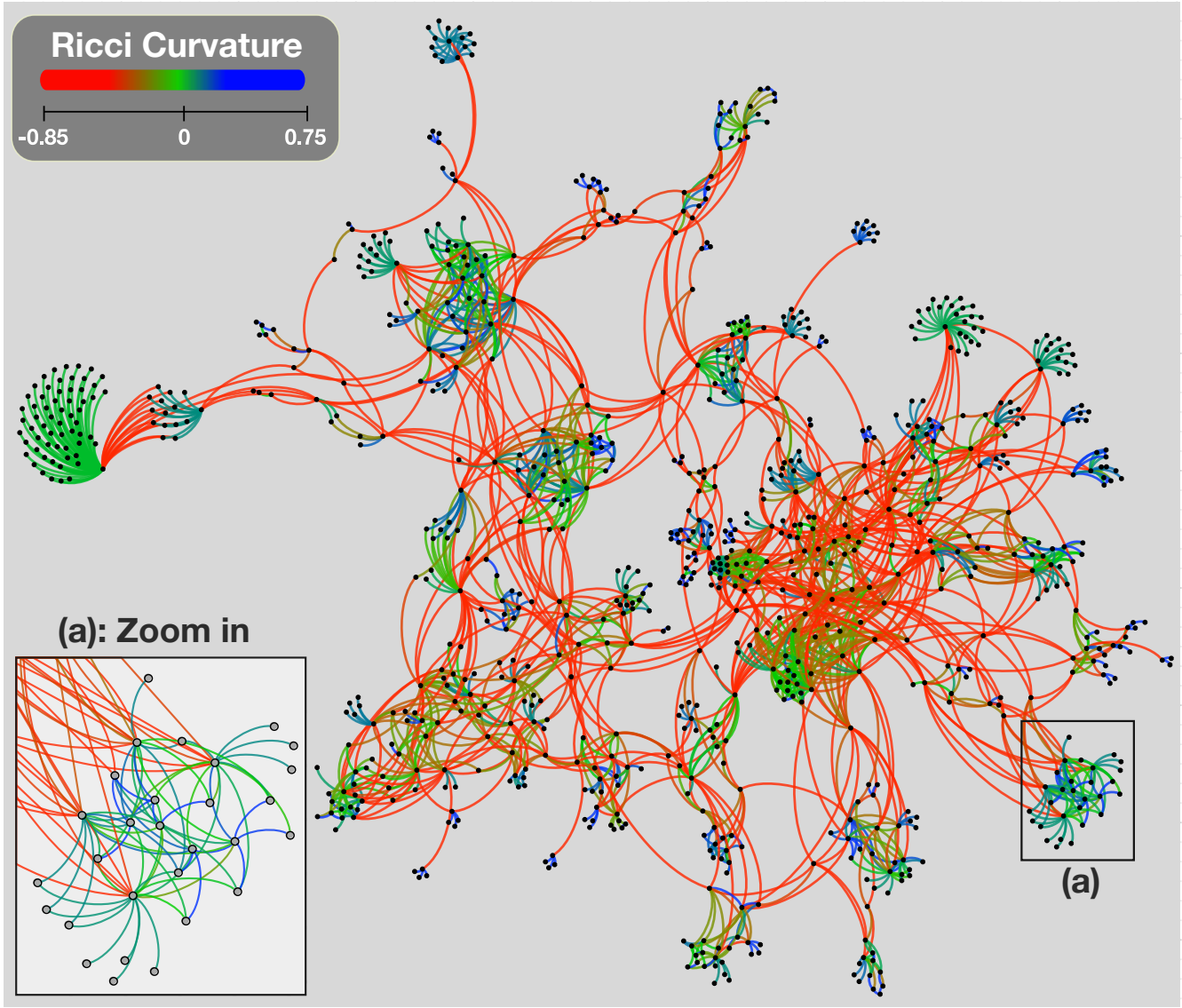

Fig. 1. It shows the Ricci curvature of each edge in a router level graph (Exodus(US)) from the Rocketfuel data set with 895 nodes and 2071 edges. Negatively curved edges (in red) behave like "backbones", maintaining the connectivity of clusters that are grouped by zero and positively curved edges (in green and blue).

**Gromov Hyperbolicity.** Recent work on measuring network traffic shows that traffic on the Internet tends to go through a relatively small core of the network as if the shortest path are attracted inwards [13]. It is then suggested that the global curvature of the network might be negative. To get some intuition, consider shortest paths for points of a hyperbolic Poincare disk, a negatively curved space. Congestion at the center is high, since geodesics tend to bend inwards.

To characterize this geometric feature, people study the Gromov hyperbolicity of such networks [14], [15]. Gromov hyperbolicity, defined on a metric space, captures the basic common features of "negatively curved" spaces like the classical real hyperbolic space $\mathbb{H}^d$, Riemannian manifolds of strictly negative sectional curvature, and of discrete spaces such as trees and the Caley graphs. In particular, a metric has $\delta$-hyperbolicity if all geodesic triangles are $\delta$-slim, that is, for any three points $a, b, c$, the geodesic paths (shortest paths) between them satisfy the following property: any point on one path is within distance $\delta$ from the closest vertex on the other two paths. See Figure 2 for an example. For a finite graph, the parameter $\delta$ is trivially finite. Thus it was suggested to compare $\delta(\triangle abc)$ with the diameter of the triangle $\triangle abc$. The ratio is bounded above by $3/2$ in a Riemannian manifold of constant nonpositive curvature, thus it was suggested to use $3/2$ as the threshold to distinguish whether the graph metric is hyperbolic or not [16].

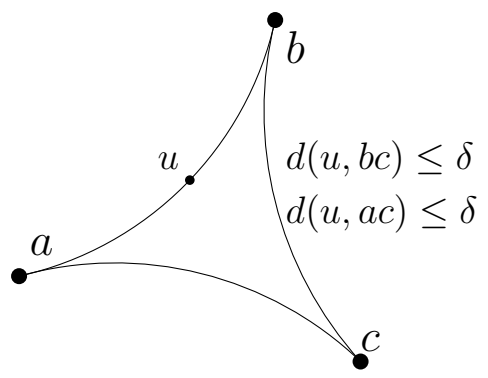

Fig. 2. Thin triangle property for a $\delta$-hyperbolic metric. The distance from any vertex $u$ on the shortest path $ab$ is within distance $\delta$ from shortest paths $bc$ and $ac$.

Study of the Rocketfuel topology [13] reveals that the absolute value of $\delta$, measured by hop count, is between $1 \sim 3$ for different ISPs. This value is much smaller than the diameter of the network (mostly in the range of $12 \sim 14$), which confirmed the initial guess of the negative curvature. For further evidence of the hyperbolicity, for each ISP topology, the highest traffic load for all pairs traffic using shortest path routing was computed, and was plotted with the size of the ISP topology. The plot shows the congestion scales in the order of ${N_i}^2$, where $N_i$ is the size of the $i$th ISP. This can be compared with the maximum congestion $n^{3/2}$ for Euclidean disks and $n^2$ in hyperbolic disks [17] (Here $n$ refers to the area of the disk), which implies the hyperbolicity of the Rocketfuel topology.

Notice that $\delta$-hyperbolicity is only concerned about the worst case metric property. What we would like to understand in this paper, is whether there is any local geometric structure (such as local curvatures) and how such geometric structures connect to network behaviors at a much finer scale.

### B. Curvature

We first review curvature in math and then describe our analysis of discrete local curvatures on Internet topology.

**Gaussian Curvature.** In mathematics, curvature is a measure of the amount by which a geometric object deviates from being flat, or straight in the case of a line, and has a number of definitions depending on the context. Most often we talk about the intrinsic curvature, which is defined at each point in a Riemannian manifold and is independent of the embedding (e.g., how the Riemannian manifold is realized in an ambient Euclidean space). Take the case of a surface $M$ in $\mathbb{R}^3$. Consider a point $p$ on the surface and a tangent vector $\mathbf{T}$ at $p$. The plane containing the tangent vector $\mathbf{T}$ and the surface normal $\mathbf{N}$ cut out a curve of the surface $M$ through $p$. The *normal curvature* is defined as the curvature of this curve at $p$. Changing the direction of the tangent vector $\mathbf{T}$ leads to different normal curvatures. The maximum and minimum normal curvature for all possible tangent vectors at $p$ are called the *principal curvatures*, $k_1$ and $k_2$, and the directions of the corresponding tangent vectors are called principal directions. The *Gaussian curvature* is the product of the principal curvatures, $k_1k_2$. Intuitively, it is positive for spheres, negative for one-sheet hyperboloids and zero for planes. It determines whether a surface is locally convex (when it is positive) or locally saddle (when it is negative).

The notion of curvature can be defined on a discrete, triangulated surface $M$ as well, see [18]. This notion of discrete curvature, however, does not work on general graphs as it requires a triangulation.

**Sectional and Ricci Curvature.** The definitions of curvatures that are easier to generalize to a discrete graph setting are sectional curvature and Ricci curvature. Consider a point $x$ on a surface $M$ and a tangent vector $v$ at $x$ whose endpoint is $y$. Take another tangent vector $w_x$ at $x$ and imagine transporting $w_x$ along vector $v$ to be a tangent vector $w_y$ at $y$. Denote the endpoints of $w_x, w_y$ as $x', y'$. If the surface is flat, then $x, y, x', y'$ would constitute a rectangle. Otherwise, the distance between $x', y'$ will differs from $|v|$. The difference can be used to define sectional curvature [1] (Figure 3).

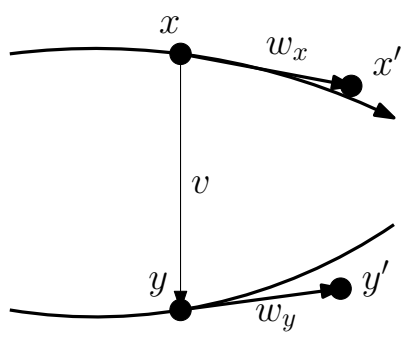

Fig. 3. Transport tangent vector $w_x$ on $x$ to be $w_y$ along vector $v$. The divergence or convergence of $w_x$ and $w_y$ are used to define sectional curvature.

The sectional curvature depends on two tangent vectors $v, w$. The Ricci curvature is obtained by averaging the sectional curvature over all directions $w$ and only depends on $v$. Intuitively, if we think of a direction $w$ at $x$ as a point on a small sphere $S_x$ centered at $x$, on average Ricci curvature controls whether the distance between a point of $S_x$ and the corresponding point of $S_y$ is smaller or larger than the distance $d(x, y)$. Ricci curvature can also be understood as representing the amount by which the volume of a geodesic ball in a curved Riemannian manifold deviates from that of the standard ball in Euclidean space. If Ricci curvature at $x$ is non-negative, the volume growth with respect to the radius of the ball centered at $x$ is a polynomial; if the Ricci curvature is negative everywhere, the volume growth is exponential.

**Discrete Ricci Curvature.** The fact that sectional curvature and Ricci curvature are mainly defined intrinsically using the geodesics on the manifold allows them to be generalized to discrete settings. Ricci curvature was introduced by Chung and Yau on graphs [19], by Ollivier [1] on Markov chains, by Bakry and Emery [20], Lott, Villani [21] and Sturm [22], and Bonciocat and Sturm [23], [24] on general metric spaces.

Ollivier's coarse Ricci curvature directly translates the definition of Ricci curvature as in Figure 3. The point is to compare the distances between two small balls with the distance between their centers. Ricci curvature is positive if the small balls are closer than their centers are. The distance between balls is defined by the well known optimal transport distance (a.k.a. Wasserstein distance, earth-mover distance, Monge-Kantorovich-Rubinstein distance).

### C. Contribution

In this paper we initiate the study of coarse Ricci curvature of the Internet topology. We would like to understand the distribution of Ricci curvature on the Internet and how the distribution varies for Internet topologies at different levels (router, AS, overlay). We want to understand the connection of Ricci curvature to other local/global graph structures (con-nectivity and vulnerability) as well as routing/path properties such as congestion.

Our observations are the following: 1) Ricci curvature can be computed fast on networks with power law degree distribution. The expected cost for each edge is a small constant, confirmed by both theory and practice. 2) In all networks, the distribution of Ricci curvature shows a large spread, implying non-homogeneity. Though each one has a fairly large fraction of negatively curved edges, there are lots of communities of positive curved edges. The curvature distribution for different type of Internet topology can be very different, suggesting that no single generative models can be used to explain all different topologies, though they share certain graph properties such as node degree distribution or small world property. 3) The Ricci curvature has a variety of connections to other graph properties (global or local), yet not predictable by anyone. This shows that the curvature measure is a new measure that adds to the richness of complex network theory. 4) For topologies representing physical connections (ISP, AS-level), edges of negative curvature more tend to stay on shortest paths in general. Removing edges of negative curvatures will disconnect network fast.

We also perform analysis of Ricci curvature on complex network models, including the Erdős-Rényi graph, the random regular graph, the power-law configuration model, the Watts Strogatz small world graph, the preferential attachment graph, and the hyperbolic grid. These model graphs share some properties of Ricci curvature with the real Internet topology but none of them was able to fully explain the real data. This suggests that the study of Ricci curvature on such complex network reveals more detailed structures than what can be captured in existing theory/models. Thus new theory is needed in this front.

## II. Ricci Curvature on Graphs

We first explain the Wasserstein distance that is used to define Ollivier's Ricci curvature.

*Definition 1 (Wasserstein distance or Earth Mover distance):* Let $X$ be a metric space with two probability measures $\mu_1$, $\mu_2$ with mass 1 respectively. A transportation plan from $\mu_1$ to $\mu_2$ is a measure $\xi$ on $X \times X$ that is mass-preserving, i.e., $\int_y d\xi(x,y) = d\mu_1(x)$ and $\int_x d\xi(x,y) = d\mu_2(y)$, where $d\xi(x,y)$ represents the amount of mass traveling from $x$ to $y$. The Wasserstein distance between $\mu_1$ and $\mu_2$, denoted by $W(\mu_1,\mu_2)$, is the minimum average traveling distance that can be achieved by any transportation plan:

$$W(\mu_1,\mu_2) = \inf_{\xi} \int \int d(x,y) d\xi(x,y).$$

The *coarse Ricci curvature* [25] considers a given metric space $(d, X)$ and a probability distribution $m_x$ on $X$, defined for each node $x$. The Ricci curvature $\kappa(x,y)$ for a pair of nodes $x,y \in X$ is obtained by comparing the earth mover distance from $m_x$ to $m_y$ and the distance $d(x,y)$:

$$\kappa(x,y) = 1 - \frac{W(m_x,m_y)}{d(x,y)}.$$

We consider a particular probability distribution $m_x$ which uses a parameter $\alpha$, as in [2], for a graph $G$. For a vertex $x \in G$ with degree $k$, let $\Gamma(x) = \{x_1, x_2, \cdots, x_k\}$ denotes the set of neighbors of $x$. For any $\alpha \in [0,1]$, the probability measure $m_x^\alpha$ is defined as

$$m_x^\alpha(x_i) = \begin{cases} \alpha & \text{if } x_i = x \\ (1-\alpha)/k & \text{if } x_i \in \Gamma(x) \\ 0 & \text{otherwise .} \end{cases}$$

The Ricci curvature as above is defined on edges. We can define the Ricci curvature of a vertex as the average of the curvature of its adjacent edges.

In the above definition we can associate any weight/metric for the input graph and compute the Ricci curvature for that metric. In our data sets, since we lack detailed information such as the delay, bandwidth, or routing policy. We take all edges of weight 1 as their hop distance beteween nodes in the computation of the Ricci curvature. This issue will be discussed in the last section.

To get an idea of Ricci curvature, here are some examples. On a square lattice graph, each edge has a zero Ricci curvature. For a tree topology, the Ricci curvature is negative for all edges except those connecting leaves. On a complete graph with at least two vertices, the Ricci curvature is all positive. In general, negative Ricci curvature means the edge behaves locally as shortcut or bridges. Positive Ricci curvature of $xy$ indicates that locally there are more triangles in the neighborhood of $x, y$.

To compute the $\alpha$-Ricci curvature, we compute the Wasserstein distance by linear programming (LP). We assume the transportation plan of $W(m_x^\alpha, m_y^\alpha)$ to be variables represented by an $m \times n$ matrix $\rho_{ij} \geq 0$, in which $\rho_{ij}$ refers to the fraction of mass transported from vertex $x_i$ to $y_j$. This transportation plan needs to preserve total mass and we are looking for the optimal plan that minimizes the total transportation distance. The LP formulation is as follows.

$$\begin{aligned} \text{Min:} \quad & \sum_j \sum_i d(x_i, y_j) \rho_{ij} m_x^\alpha(x_i) \\ \text{S.t.:} \quad & \sum_j \rho_{ij} = 1 \quad \forall i, \quad 0 \leq \rho_{ij} \leq 1 \quad \forall i, j \\ & \sum_i \rho_{ij} m_x^\alpha(x_i) = m_y^\alpha(y_j) \quad \forall j \end{aligned}$$

We notice that computation of the Ricci curvature of an edge $xy$ requires only the knowledge of the shortest path length between neighbors of $x$ to neighbors of $y$. This can be done with local knowledge. Thus the computation is scalable to large networks. We show running time analysis in Section IV.

## III. Networks and Datasets

We analyzed four types of Internet topologies: the AS-level network, router-level network, peer-to-peer file sharing network, Pretty Good Privacy (PGP) networks. As a comparison, we also analyzed topologies from power-grid network, which is known to differ from Internet topologies. The detailed characteristics are listed in Table I.

| Data Set Name | # of Node | # of Edge | Collected Time | Max. Degree | Avg. Degree | Diameter | Mean Shortest Path Length |
|---|---|---|---|---|---|---|---|
| UO Route Views Project [5], [26] | 6474 | 12572 | 2000-01 | 1458 | 3.88 | 9 | 3.71 |
| Gnutella network [27], [28] | 6301 | 20777 | 2002-08 | 97 | 6.59 | 9 | 4.64 |
| PGP network [29], [30] | 10680 | 24340 | 2001-07 | 205 | 4.56 | 24 | 7.48 |
| US power grid network [29], [31] | 4941 | 6594 | 1998 | 19 | 2.67 | 46 | 18.99 |

TABLE I
FOUR REAL-WORLD NETWORKS.

| Data Set Name | # of Node | # of Edge | Max. Degree | Avg. Degree | Diameter | Mean Shortest Path Length |
|---|---|---|---|---|---|---|
| AS: 7018, AT&T (Global) | 10145 | 14166 | 78 | 2.04 | 11 | 6.95 |
| AS: 3257, Tiscali (EU) | 843 | 1156 | 90 | 2.23 | 14 | 5.27 |
| AS: 3967, Exodus (US) | 895 | 2071 | 75 | 4.63 | 13 | 5.94 |
| AS: 1221, Telstra (Aust) | 2998 | 3789 | 106 | 2.53 | 12 | 5.53 |

TABLE II
THE ROCKETFUEL DATA SETS WITH VARIOUS COUNTRIES.

**AS Level Network.** An AS-level topology represents each autonomous system as a single node and uses a single edge to demonstrate an inter-domain connection. In this study we take AS connectivity and treat the graph as an undirected, unweighted graph. For the AS graph, we use the AS dataset from the University of Oregon Route Views Project [5], [26].

**Router Level Network.** The router level network is composed of nodes as routers and edges as the physical connections between them. Compared to the AS-level topology, the router level topology is at a finer granularity. We take the data sets from the Rocketfuel project [9], [32], see Table II. The rocketfuel data set also marks certain edges as the 'backbone edge'. It includes IP addresses and geographical locations for some nodes, that we used in the analysis.

**Peer-to-peer Networks.** In a peer-to-peer (P2P) network, computers/peers may directly connect to each other and share files directly. The connections are summarized in an overlay topology specified by the application protocol. The connections are mainly logical rather than physical. We look at such an overlay topology utilized by the Gnutella file sharing protocol. The data is collected by Ripeanu *et al.* [28] and acquired from SNAP [26].

**PGP Networks.** PGP Network characterizes the interaction of users using the Pretty Good Privacy (PGP) algorithm. It records the social behavior when people contact with each other using PGP. The initial data is collected by Boguñá *et al.* [30], the network we applied only contains the giant connected component of the network acquired from KONECT [29].

**Power Grid Networks.** As a comparison we also included a data set from the power grid network. This undirected network contains information about the power grid of the Western States of the US. Each edge represents a power supply line; each node can be any of a generator, a transformator or a substation. The power grid is typically believed to have very different structures from the Internet topology. For example, the power grid network is mostly planar and does not have a small world property. The data we use was originally used in Watts and Strogatz [31], acquired from KONECT [29].

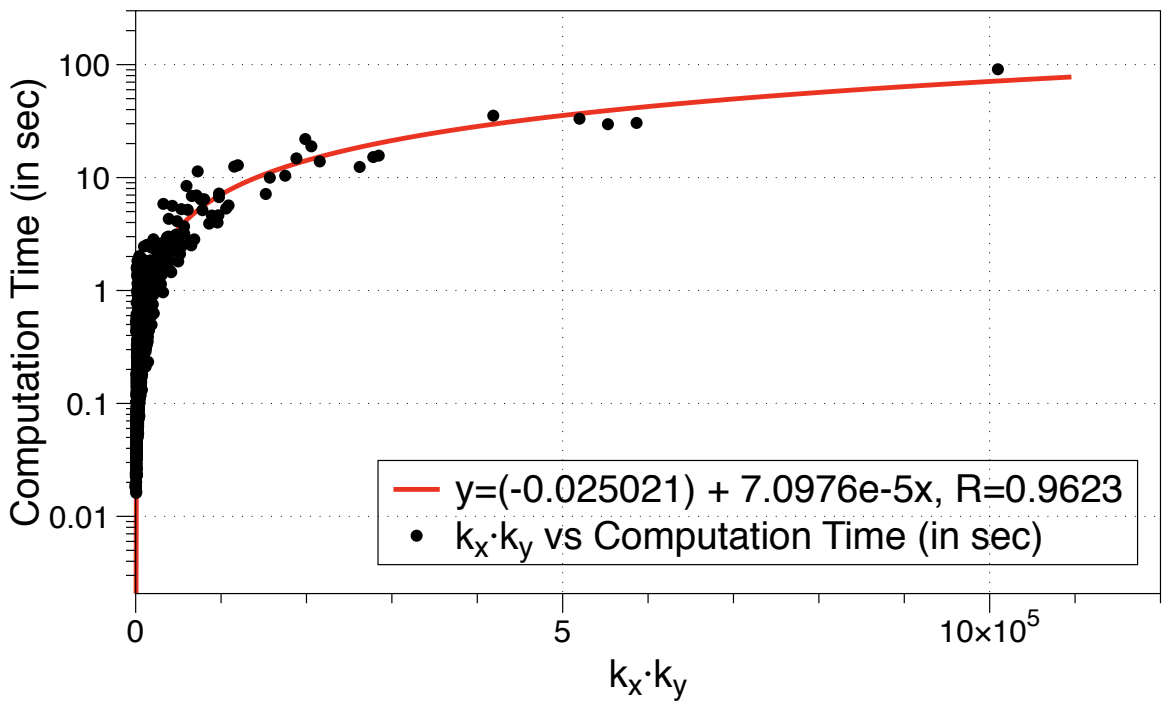

Fig. 4. An edge Ricci curvature running time comparison with the product of the edge's neighbor sizes over Route Views data set, running on Quad core 2.6 GHz Intel Core i7 with *CPLEX 12.3* [33] as LP solver. R is the correlation coefficient of the fitting curve.

## IV. RICCI CURVATURE OF THE INTERNET

### A. Running Time Analysis

Firstly, we want to understand the scalability of computing the Ricci curvature on real world topologies. Figure 4 shows the running time for computing Ricci curvature of each edge in Route Views data set. The cost is dominated by the cost of running the LP solver, which turns out to be linear in the number of variables, i.e., $k_x \cdot k_y$, for the edge $\overline{xy}$, where $k_x$ is the degree of $x$.

We can also observe that the running time cost reflects the power law degree property of the network – most edges have low degree while a few of them have very high degree. In this setting, the average running time for computing Ricci curvature for each edge is in fact a constant. Indeed, assume that the degree distribution follows a power law of exponent $\gamma \in (2,3)$, i.e., the probability of a node with degree $k$ is $\frac{\gamma-1}{k^\gamma}$. Suppose $x, y$ each chooses its degree independently from this distribution and the LP solver takes time $O(k_x k_y)$, the expected running time is

$$\int_k \int_l k\ell \cdot \text{Prob}\{k_x = k\} \cdot \text{Prob}\{k_y = \ell\} = (\frac{\gamma-1}{\gamma-2})^2 = O(1)$$

This shows that the cost of computing all Ricci curvatures in such graphs scales in the linear order of the number of edges of the network, thus suitable for large-scale networks.

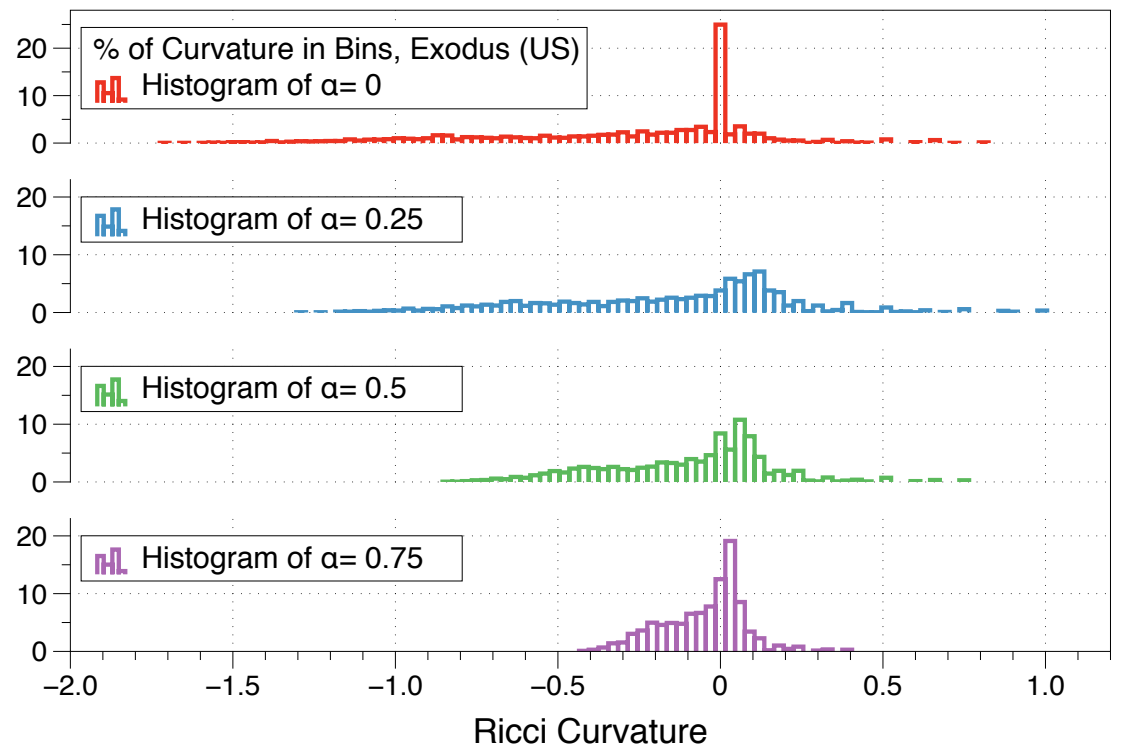

Fig. 5. A comparison of histograms of Ricci Curvature distribution with different $\alpha$ for Exodus(US) in Rocketfuel data set. Different $\alpha$ results in the same pattern of distributions curvatures, the peak of the distributions are all around $\kappa = 0$.

### B. Curvature Distribution

**The effect of $\alpha$.** We tested different values of $\alpha$ in $\alpha$-Ricci curvature. When $\alpha$ is smaller, more mass is spread to the neighbors, which intensifies the influence of neighbor nodes on curvature. Figure 5 shows the histograms of different $\alpha$ on the Exodus(US) ISP in the Rocketfuel data. Notice that the value of $\alpha$ essentially changes how the curvature distribution spreads but does not change the general shape. In the following experiments, we fix the $\alpha$ to be $0.5$.

**Curvature on various networks.** Figure 6 shows the edge curvature distribution on four networks. The curvature distribution in the AS network is similar to that of router level networks. This shows some level of self-similarity of the Internet physical topology at different resolutions. In both cases, a fraction of them have zero curvature – thus being locally 'flat', and there are more negatively curved edges than positively curved edges. In contrast, overlay networks such as Gnutella shows a heavy concentration in the negative curvature range, while in the PGP network there are more positively curved edges. It is thus intriguing to question why they generate such different behaviors. Clearly they cannot be explained by a single generative graph model even though they are similar on certain graph properties (such as power law degree etc).

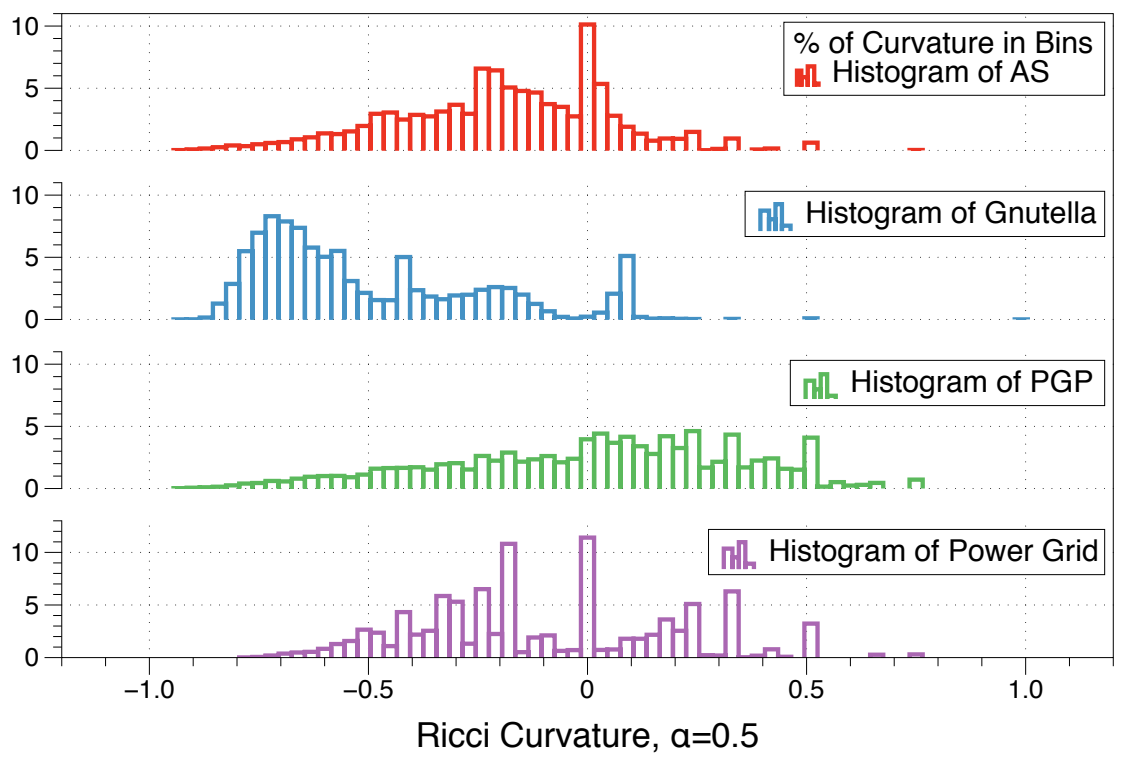

Fig. 6. A comparison of histograms of Ricci Curvature distribution in various data sets. The distribution of $\kappa$ varies with different kind of data sets.

### C. Curvature vs. Graph Connectivity

By definition, an edge $\overline{xy}$ is negatively curved meaning that locally $\overline{xy}$ behaves as a shortcut. However, we discover that in these real world graphs, the curvature of an edge is closely related to global network connectivity.

We conduct two experiments. In the first experiment, we add edges one by one, in the order of increasing curvature. We look at the number of connected components of the graph in the process. In the second experiment, we add edges in the order of decreasing curvature. Figure 7(a) shows the results. By adding edges of increasing curvature (as shown by the solid curve), the number of connected components remains small. However, adding edges with decreasing curvature (as shown by the dotted line) may generate a large number of connected components in the process. In this case, notice that the number of connected components peaks when the Ricci curvature of edges pass zero. This means that removing negatively curved edges will disconnect the entire network into small pieces, while removing all positively curved edges will not affect the connectivity that much. We also observe that all the backbone routers (as marked in the Rocketfuel data set) are negatively curved.

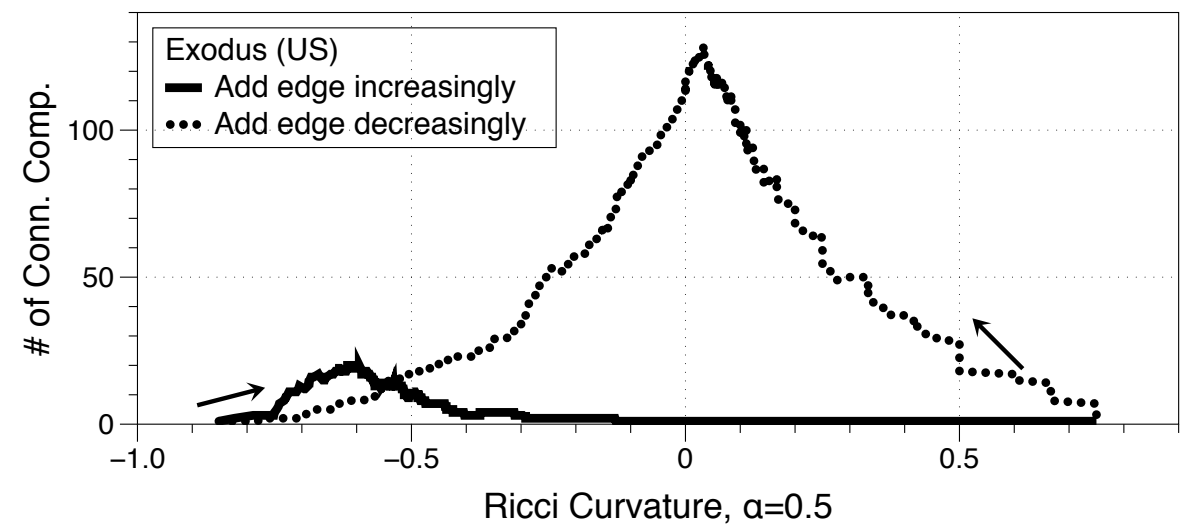

(a) Number of Connected Components

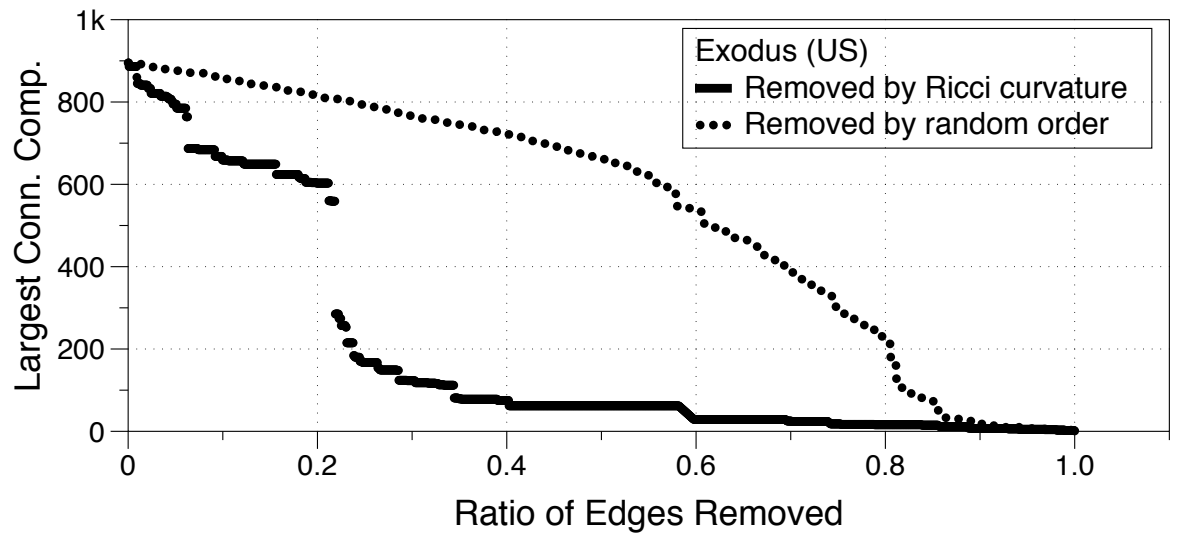

(b) Size of Largest Connected Component

Fig. 7. A comparison of number of connected components and size of largest connected component in graph for Exodus(US) in Rocketfuel data set.

Figure 8 shows the outcomes for four other types of networks with the same behavior pattern, with the only exception of Gnutella network in which the cutoff threshold seems to move much to the left.

**Robustness vs Vulnerability.** We also look at the *size of the largest connected component* when edges are removed in increasing order of their Ricci curvature. This illustrates the vulnerability of the network when the network undergoes targeted attack of removing the 'local bridges'. See the solid curve in Figure 7(b). When 20% of the most negatively curved edges are removed, the largest connected component suddenly splits into two parts. This shows as a deep drop in the curve.

As a comparison, we examine the size of the largest connected component when edges are randomly removed, shown by the black dotted line in Figure 7(b). The network connectivity holds up pretty well. The size of the largest connected component decreases steadily until most of the edges (80%) are removed. The network is fairly robust to random failures but however are vulnerable to targeted attacks on edges with negative Ricci curvature.

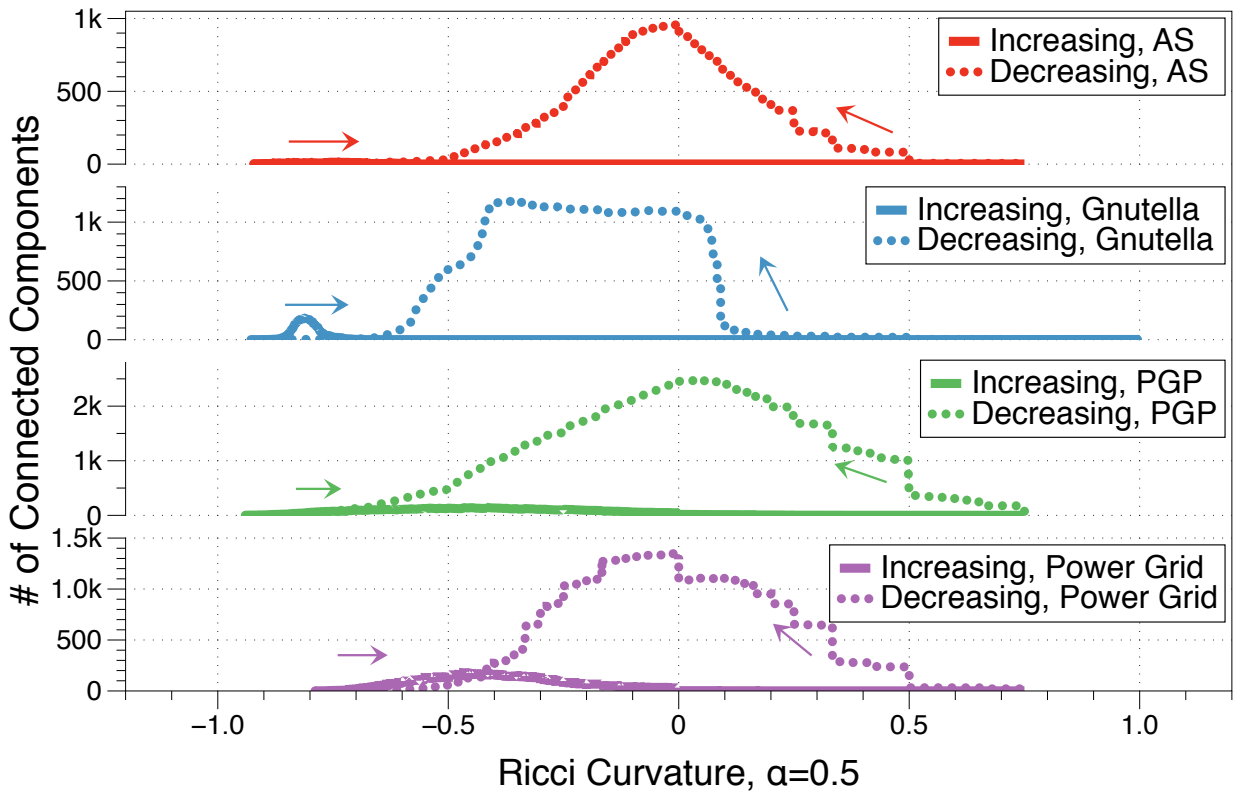

Fig. 8. A comparison of number of connected components while adding edges with curvature increasingly and decreasingly in various data sets.

## D. *Curvature vs. Global Centrality/Congestion*

**Betweenness Centrality.** The betweenness centrality [34] of an edge in a graph is the number of shortest paths that pass through that edge: $B(e) = \sum_{i,j \in G} \frac{\sigma_{ij}(e)}{\sigma_{ij}}$ where $\sigma_{ij}$ is the total number of shortest paths from node $i$ to node $j$ and $\sigma_{ij}(e)$ is the number of those paths that pass through $e$. It can be used to model congestion when shortest hop count paths are used.

Figure 10(a) and Figure 9 present the correlation between Ricci curvature and betweenness centrality. Notice that the $y$-axis is shown in exponential scale. All of five networks show a general trend that the smaller the edge curvature is, the higher betweenness centrality the edge has. However the correlation coefficient ($R$ in figures) is not really that strong, indicating that the two measures are indeed different.

**Farness Centrality.** Farness centrality of a vertex is defined as the average shortest path length to all other vertices in the same connected component. Figure 10(b) and Figure 11 show the results on five different networks. While noticing a general linear trend between farness centrality on the router level topology, the correlation is weak or none in the others.

We would like to remark that both betweenness centrality and farness centrality are global measures, whose computation requires running breadth-first search at each node and scales in the order of $O(nm)$ for a network of $n$ nodes and $m$ edges.

## E. *Curvature vs. Local Properties*

**Node degree.** Figure 10(d) and Figure 12 show curvature v.s. degree on five different networks. Although Ricci curvature is computed mainly based on the neighbor nodes, the corelation between curvature and degree is weak.

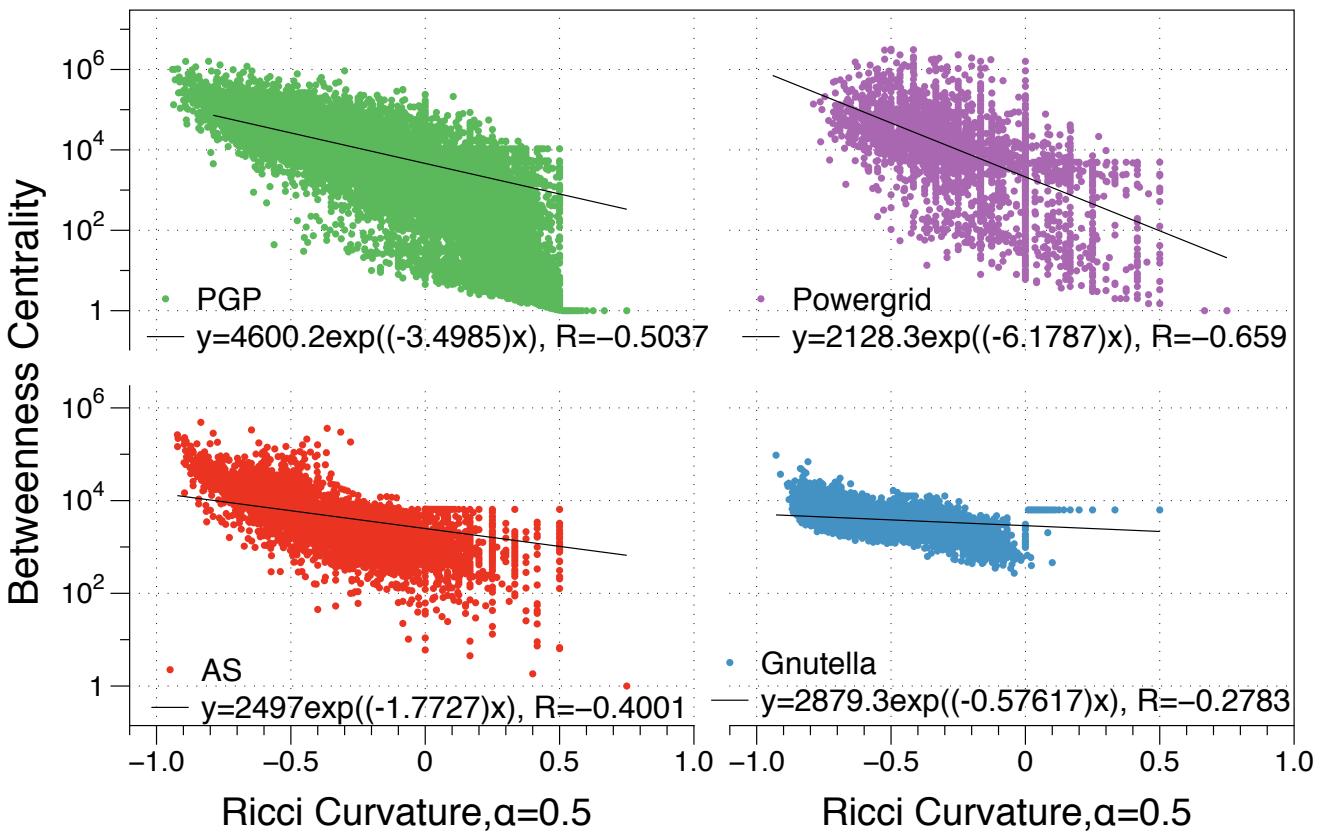

Fig. 9. A comparison of correlation between Ricci curvature and betweenness centrality on various networks.

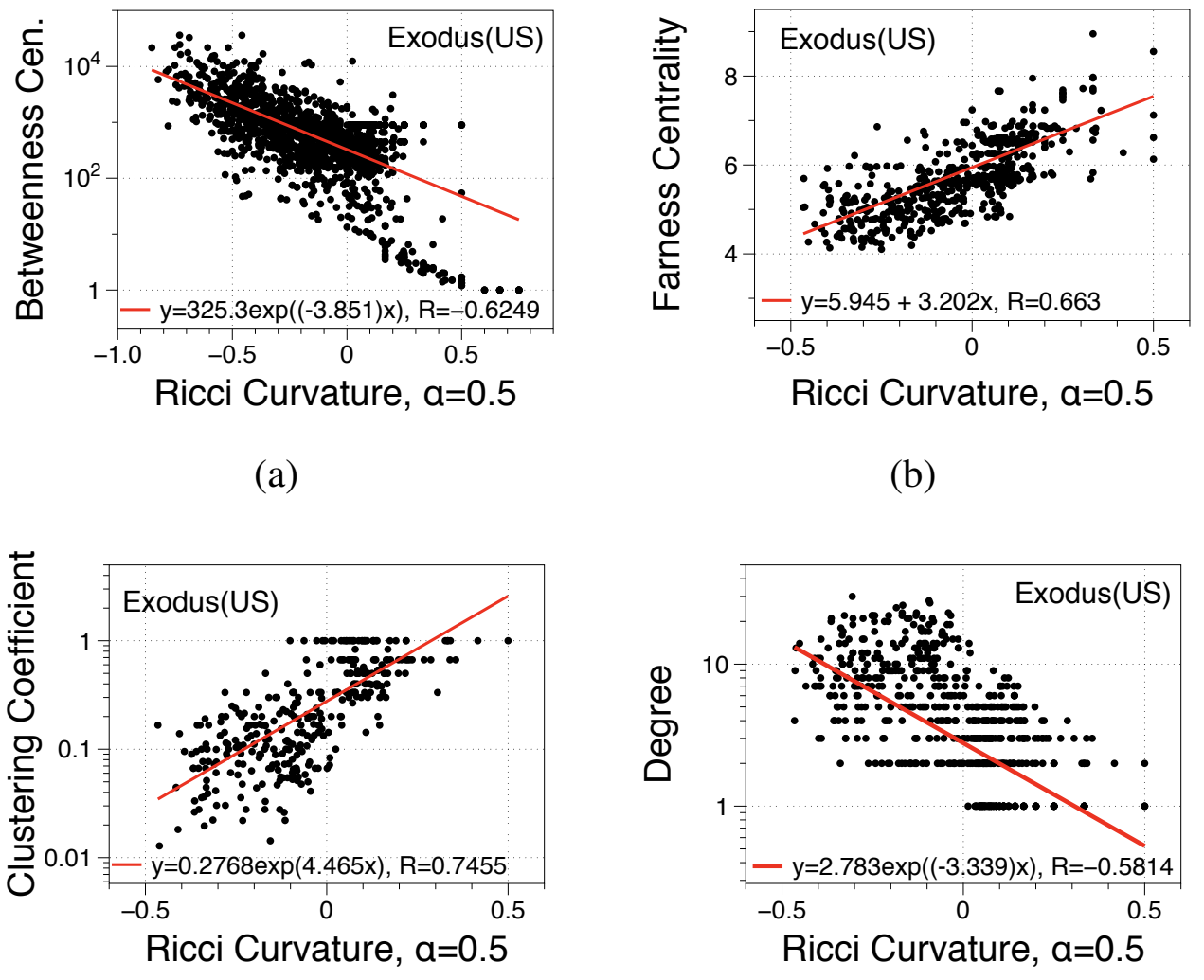

(a) (b) (c) (d)

Fig. 10. A correlation between Ricci curvature and various centralities and clustering coefficient for Exodus(US).

**Clustering Coefficient.** Clustering coefficient is a local measure capturing the extent to which vertices tend to cluster together in a graph. Specifically, in an undirected network, the clustering coefficient $C_v$ of a vertex $v$ (of degree at least 2) is defined as the number of triangles containing $v$ by the number of pairs of all neighbors of $v$, i.e. $\binom{d(v)}{2}$, where $d(v)$ is the number of neighbors of $v$ [35]. Clustering coefficient is shown to be related to the Ricci curvature in [36] (bounds are proved). But as Figure 10(c) and Figure 13 show, the correlation between clustering coefficient and Ricci curvature on nodes is rather weak.

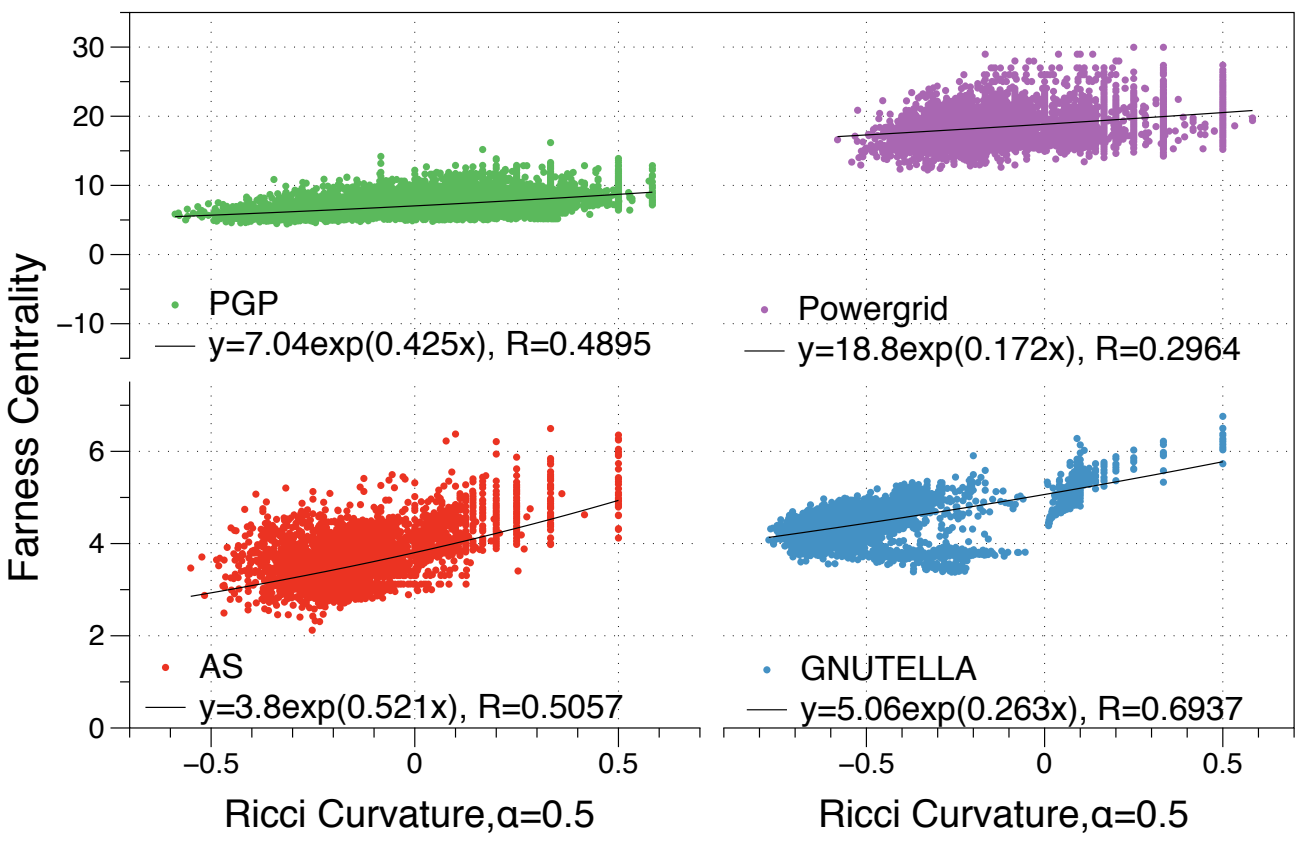

Fig. 11. A correlation between Ricci curvature and farness centrality on various networks.

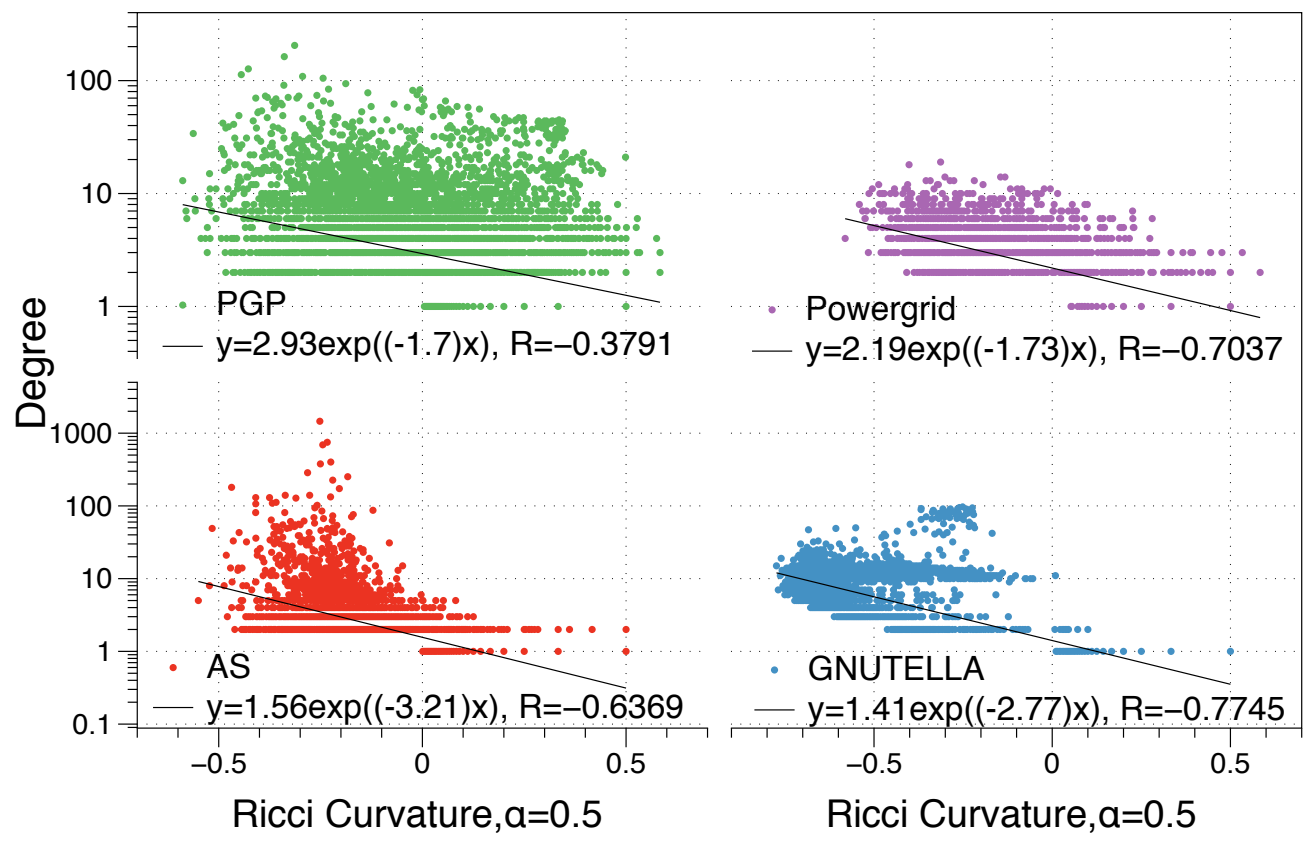

Fig. 12. A correlation between Ricci curvature and node degree on various networks.

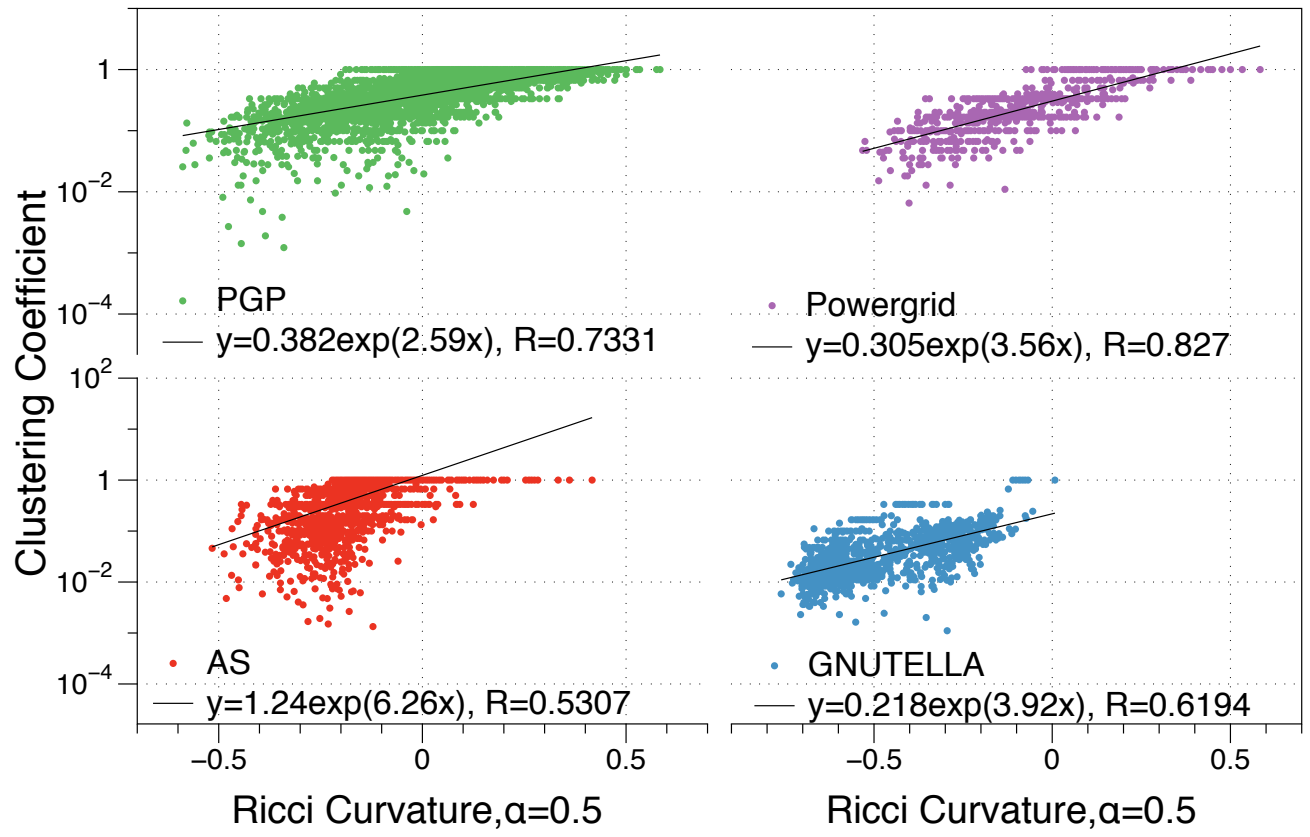

Fig. 13. A correlation between Ricci curvature and clustering coefficient on various networks.

### *F. Curvature vs. Geographical and IP Distances*

Since the Rocketfuel data sets have the geographical location and IP addresses for some of the routers, we compare them with Ricci curvature. In Figure 14, while there seems to be no special pattern for the negatively curved edges, positively curved edge tend to be geographically short. We also compare the IP distance and Ricci curvature. However, there does not seem to have any correlation.

## V. Model Networks

We examine the Ricci curvature in a number of generative models. The details of the model graphs we have tested are listed in Table III.

$G(n,p)$ **Model.** Erdős-Rényi (E-R) model [37] or $G(n,p)$ model connects each pair of vertices $\{i,j\}$ by an edge with probability $p$ independent of every other edge. We choose $(n,p)=(1000,0.01)$ in our model.

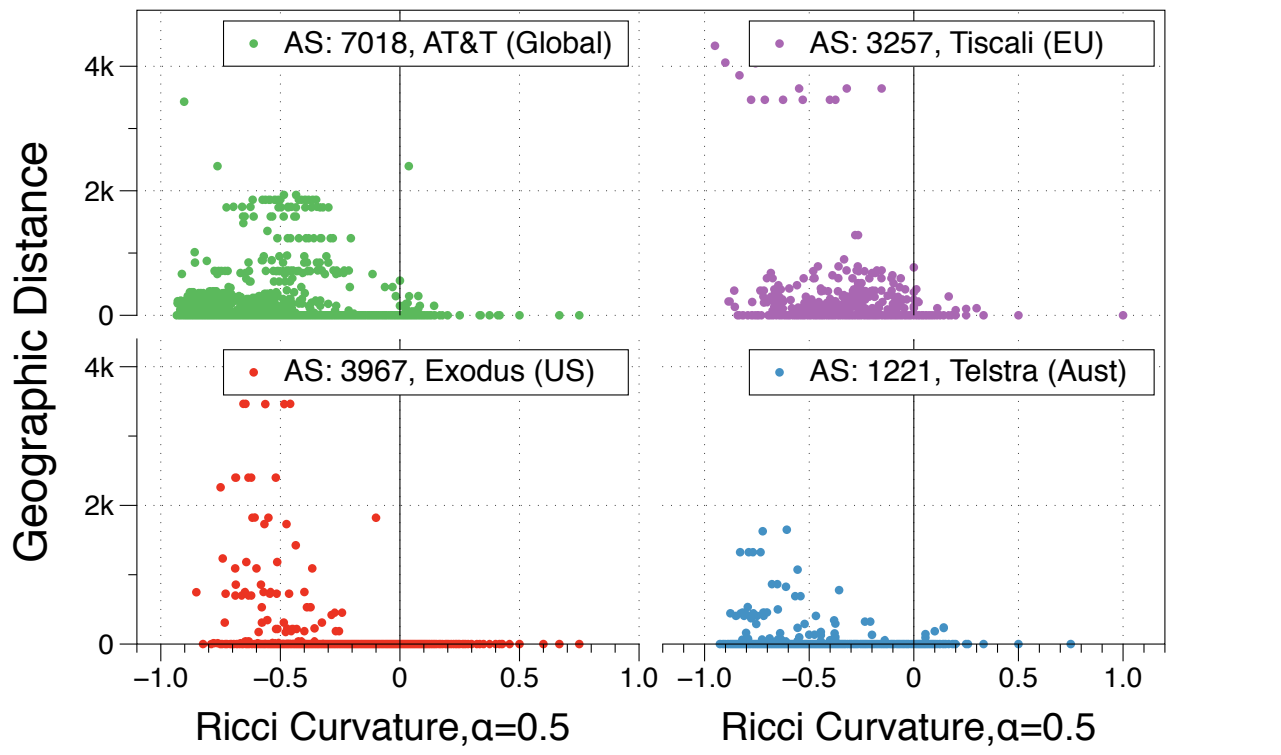

Fig. 14. A correlation between geographical distance and Ricci curvature of edges in Rocketfuel data set. We choose four different AS that belong to different countries, the results show that geographically short edges are all positively curved.

**Configuration Model.** Configuration model generates a random graph with the specified degree distribution or degree sequence. Given a degree sequence $\{d_1, d_2, \cdots, d_n\}$ where $d_i$ is the degree of vertex $i$, we give each node $i$ $d_i$ "stubs" (half edges). The random graph is constructed by a random matching of these "stubs" [38]. In the experiment, we choose the degree to be the same as the router level graph Exodus(US). This degree sequence is power law.

**Watts-Strogatz Model.** The Watts Strogatz (W-S) model [31] produces graphs with small-world properties, including short average path lengths and high clustering coefficient. The model first constructs a ring with $n$ nodes. Each node is connected to its $k$ closest neighbors. Then one randomly rewires an endpoint of each edge to a uniformly randomly chosen node with probability $\beta$. We choose $(n,k,\beta)=(1000,8,0.5)$ in our example.

**Preferential Attachment Model.** We choose Barabási-Albert (B-A) model [39]. The model starts with $k$ node initially and adds nodes one by one. Each newcomer will connect $k$ edges to existing nodes. For each edge, the endpoint is selected with probability proportional to its current degree. We take $(n,k)=(1000,2)$ in our example.

**Random Regular Graph.** A random $d-$regular graph is a graph selected uniformly from the family of all $d-$regular graph on $n$ vertices, where $3 \le d < n$ and $nd$ is even. We take $(n,d)=(1000,8)$ in our example.

**Hyperbolic Grid.** Hyperbolic grid is a tiling of the hyperbolic plane. It has negative curvature everywhere. We choose a finite subgraph of the $H(3,7)$ hyperbolic grid, which means each node in the graph is surrounded by "7" "triangles".

**Observations.** Figure 15 shows the curvature distribution of model networks. $G(n,p)$ and random regular graph have disproportionally more negatively curved edges than AS and router level topologies, and are closer to the Gnutella topology. The curvature distribution of W-S, configuration and B-A model is closer to that of AS-topology but they miss a crucial feature of a large fraction of zero curvature edges.

| Data Set Name | # of Node | # of Edge | Max. Degree | Avg. Degree | Diameter | Mean Shortest Path Length |
|---|---|---|---|---|---|---|
| G(n,p) | 1000 | 4956 | 20 | 9.91 | 6 | 3.27 |
| Watts-Strogatz | 1000 | 4000 | 14 | 8 | 6 | 3.70 |
| Random regular | 1000 | 4000 | 8 | 8 | 5 | 3.60 |
| Configuration | 895 | 2044 | 70 | 4.57 | 9 | 4.09 |
| Preferential attachment | 1000 | 1996 | 88 | 3.99 | 7 | 4.05 |
| H(3,7) | 848 | 1974 | 8 | 4.66 | 10 | 7.78 |

TABLE III
THE GRAPH PROPERTIES FOR MODEL NETWORKS.

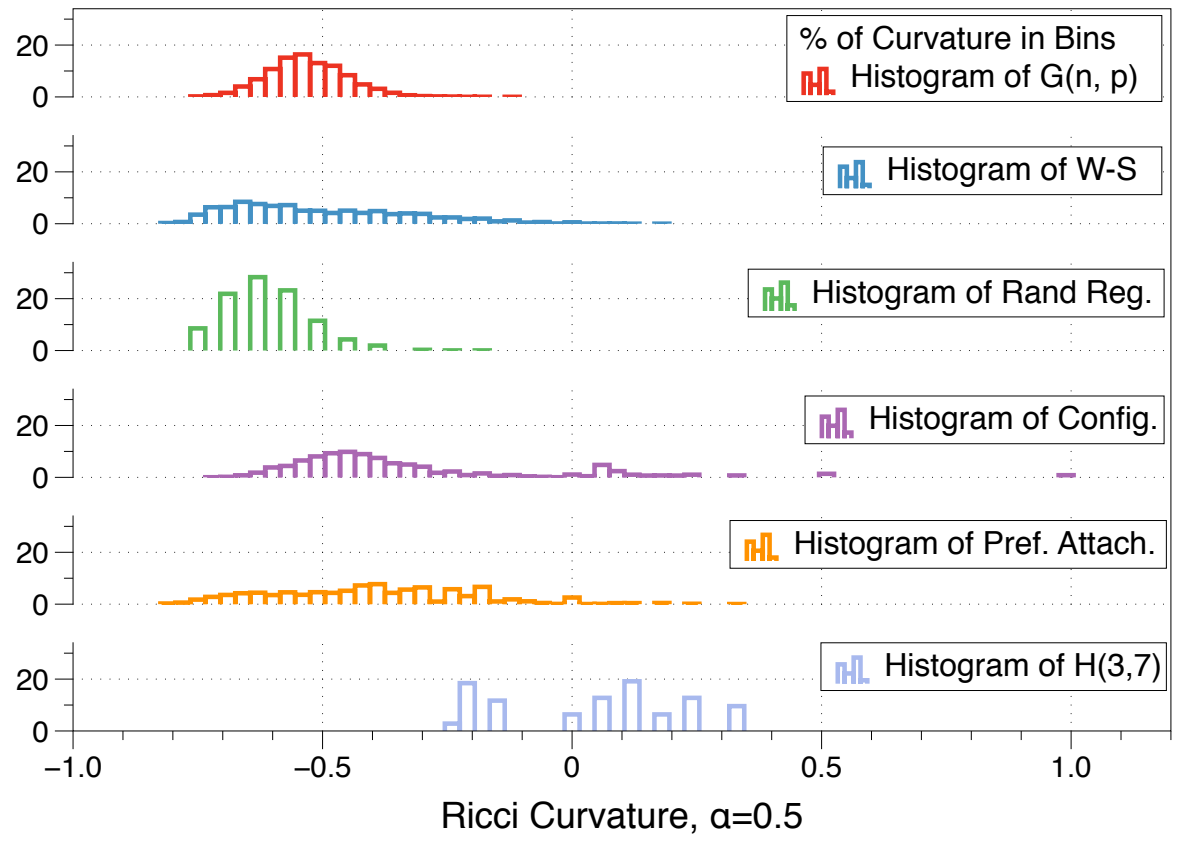

Fig. 15. A comparison of histograms of Ricci Curvature distribution in various model networks.

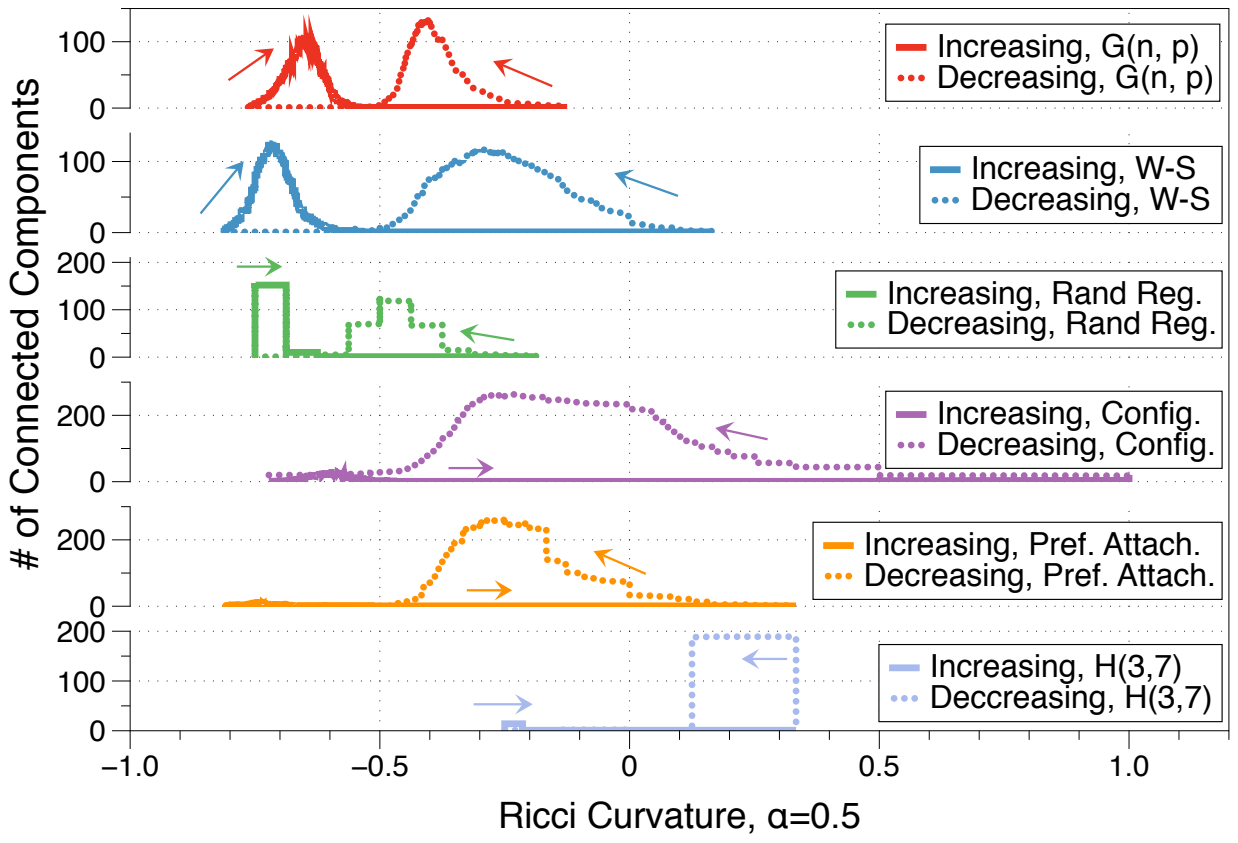

Fig. 16. A comparison of number of connected components in various model networks.

Figure 16 illustrates the results of graph connectivity experiment. In $G(n, p)$, random regular graph and W-S model, removing edges in increasing or decreasing order both create a lot of disconnected components. The models that share similar connectivity behavior to real world data are the configuration model, B-A model, and $H(3, 7)$ model. This suggests that small graph diameter has not much to do with curvature distribution/connectivity, while graph hyperbolicity and power law degree distribution are more relevant.

## VI. Discussion and Future Work

The study of Ricci curvature reveals many interesting structural properties of the Internet topology. In this section we would like to discuss the limitation of our study, the significance of the discoveries and their implications.

**Limitation of Data Sets.** We remark that our analysis is limited by the availability of data. Some of the data sets, though being almost the best available, are believed to be incomplete. For example, it is known that the AS graph is missing many non-tree like edges that connect peers not through the core. Further, the data sets we study lack important information such as the weights of the edges used in the corresponding applications (delay, bandwidth, routing policy, etc). Thus the results reported should be understood with such limitation in mind. When additional weight information is available, the Ricci curvature can be easily computed and updated. And the main goal of this paper is to start an adventure in this direction but not meant to put a conclusion to it.

**Network Embedding** Using an embedding of the network in an appropriate geometric space is one of the ingenious ideas in approximating latencies between hosts on the Internet [40]–[43]. Each host is given a coordinate that characterizes its position on the Internet such that network distances can be predicted by evaluating a distance function over hosts' coordinates. This greatly simplified the process and a well chosen set of coordinates can be highly efficient in summarizing a large amount of distance information. The embedding can also be used for selecting nearest neighbors, building a multicast tree, constructing overlay topology, and potentially any applications that require some estimates of the 'proximity' of nodes on the network [44].

The discovery that the Internet has negative curvature in the Gromov's sense is groundbreaking, since majority of the previous research that study embedding of the Internet graph use embedding into Euclidean spaces. We know from theory that a negatively curved metric cannot be embedded isometrically into an Euclidean space with fixed dimensions. Thus errors are inevitable, which may explain the relatively poor accuracy of the previous Euclidean embeddings and the improved performance of embedding in hyperbolic space [42]. Further exploration in this direction is seen in [45]–[47].

While previous work mainly use a global notion of metric curvature and only take embedding of the network into homogeneous spaces, the high curvature variation shown in this paper suggests that we may need to depart from homogeneity and instead look at a much refined granularity. We may need to think about embedding into non-homogeneous spaces such as Riemannian manifolds, which accommodate and differentiate heterogeneity expected to be present in the Internet graph.

**Internet Evolution.** One of major open questions of this work is to understand why the Internet topologies (at different levels) have the specific curvature distributions as we reported here. It would be great to develop generative graph models to explain the findings. Further, looking into the future, one can think about how to change the Internet topology to enhance its performance in terms of improving efficiency, removing vulnerability, or reducing congestion, etc. Along this direction geometric analysis may turn out to be useful. Specifically, if we formulate the desirable Ricci curvature distribution, we can examine how to change the Internet topology and link weights to achieve this ideal topology.

## Acknowledgment

Chien-Chun Ni, Yu-Yao Lin, Jie Gao and Xianfeng David Gu would like to acknowledge the founding sources NSF DMS-1418255, NSF DMS-1221339, NSF CNS-1217823, AFOSR FA9550-14-1-0193. Emil Saucan's research was supported by Israel Science Foundation Grants 221/07 and 93/11 and by European Research Council under the European Community's Seventh Framework Programme (FP7/2007-2013) / ERC grant agreement n$^{o}$ [URI-306706].

## References

[1] Y. Ollivier, "Ricci curvature of Markov chains on metric spaces," *Journal of Functional Analysis*, vol. 256, no. 3, pp. 810–864, Feb. 2009.

[2] Y. Lin, L. Lu, and S.-T. Yau, "Ricci curvature of graphs," *Tohoku Mathematical Journal*, vol. 63, no. 4, pp. 605–627, 2011.

[3] M. Faloutsos, P. Faloutsos, and C. Faloutsos, "On power-law relationships of the internet topology," *SIGCOMM Comput. Commun. Rev.*, vol. 29, no. 4, pp. 251–262, Aug. 1999.

[4] G. Siganos, M. Faloutsos, P. Faloutsos, and C. Faloutsos, "Power laws and the as-level internet topology," *IEEE/ACM Trans. Netw.*, vol. 11, no. 4, pp. 514–524, Aug. 2003.

[5] "University of Oregon route views project," http://www.routeviews.org/.

[6] S. Tauro, C. Palmer, G. Siganos, and M. Faloutsos, "A simple conceptual model for the Internet topology," 2001.

[7] M. Luckie, B. Huffaker, A. Dhamdhere, V. Giotsas, and k. claffy, "AS relationships, customer cones, and validation," in *IMC '13: Proceedings of the 2013 conference on Internet measurement conference*, Oct. 2013.

[8] Y. Shavitt and E. Shir, "DIMES: let the internet measure itself," *SIGCOMM Computer Communication Review*, vol. 35, no. 5, Oct. 2005.

[9] N. Spring, R. Mahajan, D. Wetherall, and T. Anderson, "Measuring ISP topologies with rocketfuel," *IEEE/ACM Trans. Netw.*, vol. 12, no. 1, pp. 2–16, 2004.

[10] M. Ripeanu, A. Iamnitchi, and I. Foster, "Mapping the gnutella network," *IEEE Internet Computing*, vol. 6, no. 1, pp. 50–57, Jan. 2002.

[11] J. S. Otto, M. A. Sánchez, D. R. Choffnes, F. E. Bustamante, and G. Siganos, "On blind mice and the elephant: understanding the network impact of a large distributed system," in *SIGCOMM '11: Proceedings of the ACM SIGCOMM 2011 conference*, Aug. 2011.

[12] A. Ulrich, R. Holz, P. Hauck, and G. Carle, "Investigating the openpgp web of trust," in *Computer Security ESORICS 2011*, ser. Lecture Notes in Computer Science, V. Atluri and C. Diaz, Eds. Springer Berlin Heidelberg, 2011, vol. 6879, pp. 489–507.

[13] O. Narayan and I. Saniee, "Large-scale curvature of networks," *Phys. Rev. E*, vol. 84, p. 066108, Dec 2011.

[14] M. Bonk and O. Schramm, "Embeddings of gromov hyperbolic spaces," *Geom. Funct. Anal*, vol. 10, pp. 266–306, 2000.

[15] M. Gromov, "Hyperbolic groups," in *Math. Sci. Res. Inst. Publ.* Springer, New York, 1987, vol. 8, ch. Essays in group theory, pp. 75–263.

[16] E. Jonckheere, P. Lohsoonthorn, and F. Ariaei, "Upper bound on scaled Gromov-hyperbolic $\delta$," *Applied Mathematics and Computation*, vol. 192, no. 1, pp. 191 – 204, 2007.

[17] E. A. Jonckheere, M. Lou, F. Bonahon, and Y. Baryshnikov, "Euclidean versus hyperbolic congestion in idealized versus experimental networks," *Internet Mathematics*, vol. 7, no. 1, pp. 1–27, 2011.

[18] R. Sarkar, X. Yin, J. Gao, F. Luo, and X. D. Gu, "Greedy routing with guaranteed delivery using ricci flows," in *Proc. of the 8th International Symposium on Information Processing in Sensor Networks (IPSN'09)*, April 2009, pp. 97–108.

[19] F. Chung and S. Yau, "Logarithmic harnack inequalities," *Mathematical Research Letters*, vol. 3, pp. 793–812, 1996.

[20] D. Bakry and M. Émery, "Diffusions hypercontractives," in *Séminaire de Probabilités XIX 1983/84*. Springer, 1985, pp. 177–206.

[21] J. Lott and C. Villani, "Ricci curvature for metric-measure spaces via optimal transport," *Annals of Mathematics*, vol. 169, pp. 903–991, 2009.

[22] K. T. Sturm, "On the geometry of metric measure spaces - Springer," *Acta Mathematica*, 2006.

[23] A. I. Bonciocat and K. T. Sturm, "Mass transportation and rough curvature bounds for discrete spaces," *Journal of Functional Analysis*, 2009.

[24] A. I. Bonciocat, "A rough curvature-dimension condition for metric measure spaces - Springer," *Central European Journal of Mathematics*, 2014.

[25] Y. Ollivier, "Ricci curvature of Markov chains on metric spaces," *Journal of Functional Analysis*, vol. 256, no. 3, pp. 810–864, Feb. 2009.

[26] "Stanford large network dataset collection," http://snap.stanford.edu/data/index.html.

[27] J. Leskovec, L. A. Adamic, and B. A. Huberman, "The dynamics of viral marketing," *ACM Trans. Web*, vol. 1, no. 1, May 2007.

[28] M. Ripeanu, I. Foster, and A. Iamnitchi, "Mapping the gnutella network: Properties of large-scale peer-to-peer systems and implications for system design," *IEEE Internet Computing Journal*, 2002.

[29] J. Kunegis, "KONECT – The Koblenz Network Collection," in *Proc. Int. Conf. on World Wide Web Companion*, 2013, pp. 1343–1350.

[30] M. Boguñá, R. Pastor-Satorras, A. Díaz-Guilera, and A. Arenas, "Models of social networks based on social distance attachment," *Physical Review E*, vol. 70, no. 5, p. 056122, 2004.

[31] D. J. Watts and S. H. Strogatz, "Collective dynamics of 'small-world' networks," *Nature*, pp. 440–442, June 1998.

[32] N. Spring, R. Mahajan, and D. Wetherall, "Measuring isp topologies with rocketfuel," *SIGCOMM Comput. Commun. Rev.*, vol. 32, no. 4, pp. 133–145, Aug. 2002.

[33] "IBM ILOG CPLEX optimization studio v12.3," http://www-03.ibm.com/software/products/us/en/ibmilogcpleoptistud.

[34] M. N. M. Girvan, "Community structure in social and biological networks," *Proceedings of the National Academy of Sciences USA*, vol. 99, p. 78217826, 2002.

[35] M. Latapy, "Main-memory triangle computations for very large (sparse (power-law)) graphs," *Theor. Comput. Sci.*, vol. 407, no. 1-3, pp. 458–473, Nov. 2008.

[36] J. Jost and S. Liu, "Ollivier's Ricci Curvature, Local Clustering and Curvature-Dimension Inequalities on Graphs," *Discrete & Computational Geometry*, pp. 1–23, 2011.

[37] P. Erdős and A. Rényi, "On random graphs," *Publicationes Mathematicae*, vol. 6, pp. 290–297, 1959.

[38] M. E. J. Newman, "The structure and function of complex networks," *SIAM REVIEW*, vol. 45, pp. 167–256, 2003.

[39] A. Barabási and R. Albert, "Emergence of scaling in random networks," *Science*, vol. 286, pp. 509–512, 1999.

[40] P. Francis, S. Jamin, C. Jin, Y. Jin, D. Raz, Y. Shavitt, and L. Zhang, "Idmaps: A global internet host distance estimation service," *IEEE/ACM Trans. Netw.*, vol. 9, no. 5, pp. 525–540, Oct. 2001.

[41] E. Ng and H. Zhang, "Predicting Internet network distance with coordinates-based approaches," in *Proc. IEEE INFOCOM*, 2002, pp. 170–179.

[42] Y. Shavitt and T. Tankel, "Big-bang simulation for embedding network distances in euclidean space," *IEEE/ACM Trans. Netw.*, vol. 12, no. 6, pp. 993–1006, Dec. 2004.

[43] J. Kleinberg, A. Slivkins, and T. Wexler, "Triangulation and embedding using small sets of beacons," in *Proc. 45th IEEE Symposium on Foundations of Computer Science*, 2004, pp. 444–453.

[44] Y. Shavitt and T. Tankel, "On the curvature of the internet and its usage for overlay construction and distance estimation." in *INFOCOM*, 2004.

[45] S. Lee, Z.-L. Zhang, S. Sahu, and D. Saha, "On suitability of euclidean embedding for host-based network coordinate systems," *Networking, IEEE/ACM Transactions on*, vol. 18, no. 1, pp. 27–40, 2010.

[46] F. Papadopoulos, D. Krioukov, M. Bogu, and A. Vahdat, "Greedy Forwarding in Dynamic Scale-Free Networks Embedded in Hyperbolic Metric Spaces," in *IEEE Conference on Computer Communications (INFOCOM)*. San Diego, CA: IEEE, Mar 2010.

[47] A. Cvetkovski and M. Crovella, "Hyperbolic embedding and routing for dynamic graphs," in *INFOCOM 2009, IEEE*. IEEE, 2009, pp. 1647–1655.